\documentclass[aps,prl,amsmath,twocolumn,
twoside,floatfix,superscriptaddress,preprintnumbers]{revtex4}
\usepackage{epsfig}

\def\beq{\begin{equation}}
\def\eeq{\end{equation}}
\def\bea{\begin{eqnarray}}
\def\eea{\end{eqnarray}}
\def\bq{\begin{quote}}
\def\eq{\end{quote}}
\def \lsim{\mathrel{\vcenter
     {\hbox{$<$}\nointerlineskip\hbox{$\sim$}}}}
\def \gsim{\mathrel{\vcenter
     {\hbox{$>$}\nointerlineskip\hbox{$\sim$}}}}
\def\gappeq{\mathrel{\rlap {\raise.5ex\hbox{$>$}}
{\lower.5ex\hbox{$\sim$}}}}
\def\lappeq{\mathrel{\rlap{\raise.5ex\hbox{$<$}}
{\lower.5ex\hbox{$\sim$}}}}

\def\bea{\begin{eqnarray}}   
\def\eea{\end{eqnarray}}

\begin{document}
\title{ 
 Sensitivity of the baryon
asymmetry produced by leptogenesis  to low energy CP violation
}

\author{ Sacha Davidson}
\email{s.davidson@ipnl.in2p3.fr},
\affiliation{ IPN de Lyon, Universit\'e Lyon 1, CNRS, 
 4 rue Enrico Fermi, Villeurbanne,  69622 cedex France}
\author{Julia Garayoa }
\email{julia.garayoa@ific.uv.es }
\affiliation{Depto. de F\'\i sica Te\'orica and IFIC, Universidad de 
Valencia-CSIC, Valencia,Spain}
\author{Federica Palorini}
\email{f.palorini@ipnl.in2p3.fr}
\affiliation{ IPN de Lyon, Universit\'e Lyon 1, CNRS, 
 4 rue Enrico Fermi, Villeurbanne,  69622 cedex France}
\author{Nuria Rius}
\email{nuria.rius@ific.uv.es} 
\affiliation{Depto. de F\'\i sica Te\'orica and IFIC, Universidad de 
Valencia-CSIC, Valencia,Spain}

\begin{abstract} 
If the baryon asymmetry of the Universe is produced by leptogenesis, 
CP violation is required in the lepton sector. In the seesaw extension
of the Standard  Model 
with  three hierarchical right-handed neutrinos, we show that
the baryon asymmetry is  {\it insensitive} to the PMNS phases: thermal
leptogenesis can work for any value of the observable
phases. This result was well-known when  there are no
flavour effects in leptogenesis; we show that it
remains true when flavour effects are included. 
\end{abstract}

\preprint{IFIC/07-25, 
FTUV-07-0419}

\maketitle

%%%%%%%%%%%%%%%%%%%%%%%%%%%%%%%%%%%%%%%%%%%%%%%%%%%%%%%%%%%%%%%%%%%%%%
%% INTRO        %%%%%%%%%%%%%%%%%%%%%%%%%%%%%%%%%%%%%%%%%%%%%%%%%%%%%%
%%%%%%%%%%%%%%%%%%%%%%%%%%%%%%%%%%%%%%%%%%%%%%%%%%%%%%%%%%%%%%%%%%%%%%

{\it Introduction:}
CP violation is required to  produce the puzzling excess of matter 
(baryons) over anti-matter (anti-baryons) observed in the 
Universe\cite{Sakharov:1967dj}. If this
Baryon Asymmetry of the Universe (BAU) was made via
leptogenesis \cite{review}, then 
 CP violation in the
lepton sector  is needed.
So any observation thereof, for instance in 
neutrino oscillations, 
would support leptogenesis by
demonstrating that CP is not a symmetry of the leptons. 
It is interesting to explore whether a stronger  
statement can be made about this 
 tantalising link between low-energy observable
CP violation and the  
BAU. 

In this paper, we wish to address a phenomenological question:
``is the baryon asymmetry {\it sensitive} to the phases  of the lepton
mixing matrix (PMNS matrix)? ''. 
Electroweak precision data was said to be 
sensitive to the top mass, meaning that
a preferred range for $m_t$ could be extracted from the data.
Here, we wish to ask a similar question, assuming the baryon asymmetry
is generated, via leptogenesis,   
from the decay of the lightest
``right-handed'' (RH) neutrino:  given the
measured value of the baryon asymmetry,
 can an allowed
range for the PMNS phases be obtained?

It was shown in \cite{Branco:2001pq}
that the BAU produced by thermal leptogenesis
in the type 1 seesaw, 
without ``flavour effects'',  is insensitive to PMNS phases.
That is, the PMNS phases can be zero while leptogenesis works, 
and the  CP asymmetry of leptogenesis   
can vanish for  
arbitrary values of the PMNS phases.  
In fact, the ``unflavoured'' asymmetry is
controlled by phases from the RH sector only,
and it would vanish were this sector CP conserving.
However, it
was recently realised that lepton flavour
matters in leptogenesis\cite{issues,Netal,matters}:
in the relevant  temperature range $10^{9} \rightarrow 10^{12}$ GeV,
the final baryon asymmetry depends separately on the lepton  asymmetry
 in $\tau$s    ,  and on the lepton asymmetry
in  muons and electrons. 
So in this paper,  we revisit the question addressed
in \cite{Branco:2001pq}, but with the inclusion of flavour effects.
Our analysis differs from recent  discussions
\cite{Fujihara:2005pv} (2RHN model), \cite{Pascoli:2006ie,Branco:2006ce}
(CP as a symmetry of the $N$ sector),
\cite{akr} (sequential $N$ dominance)  
in that we wish to do a bottom-up analysis of the
three generation seesaw.  Ideally, we wish  to 
express the  baryon asymmetry  in
terms of observables, such as the light neutrino
masses and PMNS matrix,  and   free parameters.
Then, by inspection, one could determine whether
fixing the baryon asymmetry constrained the PMNS phases.

{\it  Notation and review:}
We consider a seesaw model \cite{seesaw}, where
three heavy ($M \gsim 10^9 $ GeV) majorana neutrinos $N_I$ are added to
the Standard Model.  The Lagrangian at the $N_I$ mass
scale is 
\beq
{\cal L} =   \overline{e_R}^j {\bf  Y_e}_{ij}  H_d   \ell^i + 
   \overline{N}^J  {\bf  \lambda}_{iJ}  H_u  \ell^i  + 
 \overline{N}^J \frac{{\bf  M}_{JK}}{2} N^{cK} + h.c.
\label{Lag}
\eeq 
where  the 
flavour index order on the Yukawa matrices 
${\bf  Y_e}, \lambda$  is left-right, and 
 $H_u = i \sigma_2 H_d^*$.

There are 6 phases among the 21 parameters of
this  Lagrangian. 
We can work 
in the mass eigenstate basis of the charged leptons and the $N_I$,
and write the neutrino Yukawa  matrix  as 
\beq
\lambda = V_L^\dagger D_\lambda V_R ~~~~, 
\eeq
where $D_\lambda$ is real and diagonal, and $V_L, V_R$ 
are unitary matrices, each containing three phases.
So at the high scale, one can distinguish CP violation
in the left-handed doublet sector (phases that appear in $V_L$) 
and in the right-handed singlet sector (phases in $V_R$).
Leptogenesis can work when there are phases in either
or both sectors.

At  energies accessible to experiment,
well below the $N_I$ mass scale, 
 the light (LH)
neutrinos acquire an effective Majorana mass matrix
\footnote{which appears in the Lagrangian as  
$\frac{1}{2} [m \,]_{\alpha \beta} \nu^\alpha \nu^\beta + h.c$}:
\beq
\label{kappa}
  [m \,]    =   {\lambda M^{-1} \lambda^T}v^2  
= {U  D_m U^T} 
\eeq
where  $v = 174$ GeV is the Higgs vev,
$D_m$ is diagonal with real eigenvalues, and
$U$ is the PMNS matrix.
 There are nine
parameters in $[m \,]$,  which is 
``in principle'' experimentally accessible.
 Two mass differences
and two angles of $U$ are measured, leaving the mass scale, one
angle and three phases of $U$ unknown.

 From the above we can write
\bea
D_m  & = &U^\dagger V_L^\dagger D_\lambda V_R D_M^{-1}V_R^T D_\lambda
V_L^* U^* v^2
\eea
so we see that the PMNS matrix will generically have phases
 if $V_L$ and/or $V_R$ are complex.
Like leptogenesis,
it  receives contributions from CP violation in the
 LH and RH sectors. 
Thus it seems  ``probable'',  or even ``natural'',
that there is some relation between the CP violation
of leptogenesis and  of the PMNS matrix. However,
the notion of  relation or dependence    is nebulous
\cite{ss25}, so  we address  the more clear
and simple question of whether the baryon asymmetry 
is {\it sensitive} to  PMNS phases. By this we mean: if the
total baryon asymmetry is fixed, and we assume
to know all the neutrino masses and mixing angles,
can we predict ranges for the PMNS phases?

We suppose that the baryon asymmetry is made via leptogenesis,
in the decay of the lightest singlet $N_1$,  with 
$ M_1 \sim 10^{10}$ GeV.
Flavour effects
are relevant in this temperature range \cite{issues,Netal,matters},
\footnote{provided the  decay rate of $N_1$ is slower 
than the interactions of the $\tau$ Yukawa \cite{BdBR}}. 
$N_1$ decays to leptons $\ell_\alpha$, an amount $ \epsilon_{\alpha \alpha}$
 more  than  to anti-leptons $\overline{\ell}_\alpha$,
and this lepton asymmetry is transformed to a baryon
asymmetry by SM processes (sphalerons).
We will further suppose that the partial  decay rates  of $N_1$ to
each  flavour are faster than  the expansion rate 
of the Universe $H$. This implies that $N_1$
decays are close to equilibrium, and there is a
significant washout of the lepton asymmetry due
to $N_1$ interactions 
(strong washout regime); we discuss later why 
this assumption does not affect our conclusions. 

  Flavour effects are relevant in
leptogenesis\cite{issues,Netal,matters} because the final  asymmetry
cares which leptons $\ell$ are distinguishable.
$N_1$ interacts only via  its Yukawa
coupling, which controls its  production and destruction.
The washout of the asymmetry, by  decays, inverse
decays and scatterings of $N_1$, is therefore crucial
for leptogenesis to work, because otherwise the
opposite sign asymmetry generated at early times during $N_1$ production 
would cancel the asymmetry produced as they disappear.
To obtain the washout rates (for instance,
for $\ell + H_u \rightarrow N_1$), one must know the initial
state particles, that is, which leptons are distinguishable.

At $ T \sim M_1$, when the asymmetry is generated,
SM interactions can be categorised as much faster than
$H$, of order  $H$, or much slower. 
Interactions that are  slower than $H$ 
can be neglected. $H^{-1}$ is the age of the Universe
and the timescale of leptogenesis, so the  faster interactions
 should be resummed--- for instance
into thermal masses. In the temperature range
$10^9 \lsim T \lsim 10^{12}$ GeV, interactions of
the $\tau$ Yukawa are faster than $H$, so 
the $\ell_\tau$ doublet is distinguishable 
(has a different ``thermal mass'') from 
the other  two lepton doublets. 
The decay of $N_1$ therefore produces
asymmetries in $B/3 - L_\tau$, and in 
 $B/3 - L_o$, where $\ell^o$  (``other'') is
the  projection in  $\ell^e$ and $\ell^\mu$ space,
of the direction into which $N_1$ decays\cite{Barbieri}:
$\hat{\ell}_o = (\lambda_{\mu 1} \hat{\mu} + \lambda_{e 1} \hat{e})/
{\sqrt{ | \lambda_{\mu 1}|^2 + | \lambda_{e 1}|^2}} $.
 Following \cite{matters},
we approximate these asymmetries to evolve independently.
%\footnote{This is not quite the case, because 
% sphalerons ensure  
%$Y_B \propto$ the doublet lepton asymmetry
%in all flavours.
%%   $ Y_{e e} + Y_{ \mu \mu} + Y_{\tau \tau}$.  
%So $Y_{B/3 - L_o}$  large, could  contribute
%to   the $\ell_{ \tau}$  asymmetry which
%controls  the washout  rate of  $Y_{B/3 - L_\tau}$. 
%This is described by the off-diagonal  elements
%of the A-matrix  of \cite{Barbieri}; see \cite{Netal,AAFX}
%for a discussion. }.
In this case,
the baryon to entropy ratio can be written as the sum over flavour
of the flavoured CP asymmetries $ \epsilon_{\alpha \alpha}$ 
 times   a flavour-dependent washout parameter
  $\eta_{\alpha} <1$ which is obtained by solving the relevant 
flavoured Boltzmann equations \cite{issues,Netal,matters}: 
\beq
Y_B \simeq  \frac{12}{37} \frac{1}{3 g_*} \left( \epsilon_{\tau \tau} 
\eta_{\tau}
 +  \epsilon_{o o}
\eta_{o}
\right)
\label{YB}
\eeq
where $g_* =106.75 $ counts entropy, and the 12/37  is
the fraction of a $B-L$ asymmetry which,
in the presence of sphalerons,   is
stored in baryons.

In the limit of hierarchical RH neutrinos,
the CP asymmetry in the decay $N_1 \rightarrow \ell_\alpha  H$
can be written as
\bea
\epsilon_{\alpha \alpha} 
& \simeq & 
- \frac{3 M_1}{16 \pi v^2  [\lambda^\dagger \lambda]_{11}} 
{ {\rm Im}} \{ [\lambda]_{\alpha 1} [m^\dagger 
\lambda]_{\alpha 1} \} 
\label{epsaa}
\eea
where $m$ is defined in eqn (\ref{kappa}).

In the  case of  `` strong washout''  for all flavours,
 which corresponds to $\Gamma( N_1 \rightarrow \ell_\alpha H_u) >H_{(T=M_1)}$
for $\alpha = \tau, o$, 
the washout factor is approximately \cite{matters,AAFX}
\beq
\eta_{\alpha} \simeq   1.3 \left( \frac{m_*}{6 A_{\alpha \alpha}
\widetilde{m}_{\alpha \alpha}} \right)^{1.16} \rightarrow
\frac{m_*}{5 A_{\alpha \alpha}
\widetilde{m}_{\alpha \alpha}}
\label{etaa}
\eeq
where there is  no sum on $\alpha$,
 $m_* \simeq  10^{-3}$ eV, and
  $A_{ \tau \tau} \simeq A_{o o} \sim 2/3$ \cite{Barbieri,matters}
\footnote{ The A matrix parametrises the redistribution
of asymmetries in chemical equilibrium.}.
The (rescaled) $N_1$ decay rate  is
\beq
\widetilde{m} = \sum_\alpha \widetilde{m}_{\alpha \alpha}
% = \widetilde{m}_{\alpha \alpha} \Huv^2 =
%\sum_\alpha  8 \pi \frac{\Gamma( N_1 \rightarrow \ell_\alpha H)}{M_1^2} \Huv^2
=  \sum_\alpha \frac{|\lambda_{\alpha 1}|^2}{M_1} v^2
\eeq

{\it An equation:}
Combining equations (\ref{YB}), (\ref{epsaa}), and  (\ref{etaa}), we obtain 
$Y_B  \propto  { \epsilon_{\tau \tau}} / { \widetilde{m}_{\tau \tau}}
 +  { \epsilon_{o o}} / {\widetilde{m}_{oo}} 
$, 
where ($\alpha$ not summed)
\bea
\frac{ \epsilon_{\alpha \alpha}}{ \widetilde{m}_{\alpha \alpha }}
& =&
\frac{3M_1 }{16 \pi v^2  \widetilde{m}}  \sum_\beta 
{ {\rm Im}} \{ \hat{\lambda}_\alpha  m_{\alpha \beta} 
\hat{\lambda}_{\beta} \} \frac{|\lambda_\beta|}{|\lambda_\alpha|}
\eea
and the Yukawa couplings of $N_1$  have been  written
as a phase factor times a magnitude : $ \hat{\lambda}_\alpha
 | \lambda_{\alpha }| 
= \lambda_{\alpha 1}^* $.  So the  baryon asymmetry can
be approximated as
\bea
Y_B  &\simeq&  Y_B^{bd}
\left(   \frac{{ {\rm Im}} \{ \hat{\lambda}_\tau \! \!  \cdot m \! \! \cdot 
\hat{\lambda}_{\tau} \} }{m_{atm}} +
 \frac{{ {\rm Im}} \{ \hat{\lambda}_o  \! \!   \cdot m \! \! \cdot  
\hat{\lambda}_{o} \} }{m_{atm}} \right.  \nonumber \\
&&+ \left. 
 \frac{{ {\rm Im}} \{ \hat{\lambda}_\tau \! \!   \cdot m\! \! \cdot 
\hat{\lambda}_{o} \} }{m_{atm}}\left[ \frac{|\lambda_o|}{|\lambda_\tau|} +
 \frac{|\lambda_\tau|}{|\lambda_o|} \right] \right) \frac{1}{A_{\tau \tau}}
\label{belle}
\eea
The prefactor of the parentheses
$\displaystyle{ Y_B^{bd} =  \frac{12}{37} \frac{M_1 m_{atm}}{16 \pi v^2 }
\frac{ m_*}{5 g_* \widetilde{m}}}  $
 is the upper bound
on the baryon asymmetry,  that would be obtained in the
strong washout case  by neglecting flavour
effects. Recall that this equation is only valid in strong washout
for all flavours.

This equation reproduces the observation \cite{matters},
that: {\it (i)} for equal asymmetries and equal decay rates
of all distinguishable flavours, flavour effects
increase the upper bound on the  
baryon asymmetry 
by $\sum_a A_{aa}^{-1} \sim 3$. 
%\footnote{
%$A_{aa} \simeq 2/3 [1]$ in the two [three] distinguishable
%flavour case.}.
{\it (ii)}
More interestingly,  having stronger
washout in one flavour, can increase  the baryon asymmetry
[via the term in brackets]. 
So models in which the Yukawa coupling $\lambda_{\tau 1}$
is significantly different from $\lambda_{\mu 1},\lambda_{e 1} $,
can have an enhanced baryon asymmetry (with
cooperation from the phases).

Finally, this equation is attractive  step towards
writing the baryon asymmetry as a real function
of real parameters ( 
$Y_B^{bd}$, depending on  $M_1$ and $\tilde{m}_1$),
times a phase factor \cite{HMY}. In this case,
the phase factor is a sum of three terms, depending on
the phases of the $N_1$  Yukawa couplings,  light
neutrino mass matrix elements normalised by the heaviest mass,
and a (real)  ratio of Yukawas.

{\it CP violation:}
In this section, we would like to use eqn (\ref{belle}) to
show that  the baryon asymmetry is  insensitive to
the PMNS phases.  The parameters of
the lepton sector  can be  divided into ``measurables'', which
are the neutrino and charged lepton masses, and
the three angles and three phases of the PMNS matrix $U$.
The remaining 9 parameters are unmeasurable. 
We want to show that for any value of
the PMNS phases, there is at least one  point
in the parameter space of the   unmeasurables
where  a large enough
baryon asymmetry is obtained.
The approximations leading to eqn (\ref{belle}) are only
valid in a subset of the  unmeasurable parameter space,
but if we can find points in this subspace,
we are done. We first  show analytically that such points exist,
then  we do a parameter
space scan to confirm that leptogenesis can
work for any value of the PMNS phases.

If the phases of the $\lambda_{\alpha 1}$
were independent of the PMNS phases, and a big enough
$Y_B$ could be obtained for some value of
the PMNS phases,  then our claim is
true by inspection:  for any other values,  
the phases
of the $\lambda_{\alpha 1}$ could be chosen
to reproduce the same $Y_B$. However,
there is in general some relation between the phases
of $m$ and those of $\lambda_{\alpha 1}$, so
we proceed by looking for an area of parameter
space where the phases of the  $\lambda_{\alpha 1}$
can be freely varied without affecting the
``measurables''. Then  we check that a large
enough baryon asymmetry can be obtained.

Such an area of parameter space can be
found  using the  $R$ matrix parametrisation
of Casas-Ibarra \cite{Casas:2001sr}, where  the complex
orthogonal matrix $R$  is defined
such that $ \lambda v \equiv U D_m^{1/2} R D_M^{1/2}$.
Taking a simple $R$ of the form
\beq
R = \left[ \begin{array}{ccc}
\cos \phi & 0 & -\sin \phi \\
0 & 1 & 0 \\
 \sin \phi  & 0 &\cos \phi
\end{array} \right]
\eeq
and parametrising $U = VP$,
where $V$ is a CKM-like  unitary matrix
with one  ``Dirac'' phase $e^{-i \delta}$  appearing
with $\sin \theta_{13}$, and $P = {\rm diag} \{ e^{i \varphi_1/2}, 
e^{i \varphi_2/2},1 \} $, gives
\bea
\frac{\lambda_{\tau 1}v}
{ \sqrt{M_1 m_{3}}}& = & U_{\tau 1} \sqrt{\frac{m_1}{m_3}}
\cos \phi + U_{\tau 3}\sin \phi 
\simeq  \frac{\sin \phi}{\sqrt{2}}  \\
\frac{\lambda_{\mu 1}v}{ \sqrt{M_1 m_3}}& = 
& U_{\mu 1} \sqrt{\frac{m_1}{m_3}}
\cos \phi + U_{\mu 3}\sin \phi 
\simeq   \frac{\sin \phi}{\sqrt{2}}  \\
\frac{\lambda_{e 1}v}{ \sqrt{M_1 m_3} }& = &U_{e 1} \sqrt{\frac{m_1}{m_3}}
\cos \phi + U_{e 3}\sin \phi 
\eea
where we took hierarchical neutrino masses.
We neglect $\lambda_{e 1}$ because its absolute value
is small.
With this choice of the unknown  $R$, the phases of
the $\lambda_{\alpha 1}$ are effectively independent
of the PMNS phases. So for any choice of PMNS phases
that would appear on the $m$ of eqn (\ref{belle}),
the phases of the Yukawa couplings can be chosen
independently, to ensure enough CP violation for
leptogenesis.

We now check that a large enough baryon asymmetry
can be obtained in this area of parameter space.
The parentheses of eqn (\ref{belle}) can be written explicitly
as
\bea
{ {\rm Im}} \left\{ \frac{\sin^2 \phi^*}{|\sin \phi|^2} ( m_{\tau \tau}
+  m_{\mu \mu}
+ 2  m_{\mu \tau} ) \right\}
\frac{1}{m_{atm}} %\nonumber \\
\eea
Writing $\phi^* = \rho -i \omega$, 
the final baryon asymmetry can be estimated from eqn (\ref{belle}) as
\beq
\frac{Y_B}{ 10^{-10}}  \simeq   -  
\left( \frac{M_1}{10^{11}  {\rm GeV}} \right)
 \frac{\sin \rho \cos \rho \sinh \omega \cosh \omega}{
(\sin^2 \rho \cosh^2 \omega +
\cos^2 \rho \sinh^2 \omega )^2}
\eeq
which can equal the observed $8.7   ^{+0.3}_{-0.4} \times 10^{-11}$
\cite{Spergel:2006hy} for $M_1 \sim $
few $ \times 10^{10}$ GeV, and judicious choices of $\rho$ and $\omega$.

A similar argument can be made if the
light neutrino mass spectrum is inverse hierarchical.

\begin{figure}[h]%%%%%%%%%%%%%%%%%%%%%%%%%%%%%%%%%%%%%%%%%%%
  %%%
\epsfig{figure=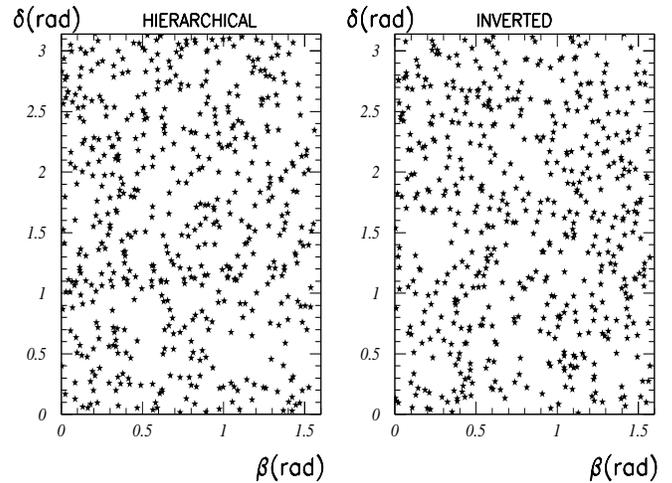,width=.55\textwidth,height=.31\textheight}
 \caption{\small A random selection of points
where the baryon asymmetry   is large enough, 
for some  choice of the unmeasurable parameters
of the seesaw.
The light neutrino masses are taken non-degenerate,  
and the Majorana  phase of
the smallest one can be neglected.
 The ``Dirac'' phase
$\delta$ is defined  such that  
$U_{e3} = \sin \theta_{13}e ^{-i \delta}$, and
$\beta$ is the  majorana phase of
$m_2 = |m_2| e^{2i \beta}$.
The baryon asymmetry arises in the decay of $N_1$
of mass $M_1 = 10^{10}$ GeV.
}
  \label{fig1}
\end{figure}

The scatter plots of figure \ref{fig1}    show
that a large enough baryon asymmetry can be obtained for
any value of the PMNS phases. 

%%%%%%%%%%%%%%%%%%%%%%%%%%%%%%%%%%%%%%%%%%%%%%%%%%%%%%%%%%%
%%%%%%%%%%%%%%%%%%%%%%%%%%%%%%%%%%%%%%%%%%%%%%%%

The plots are obtained by fixing $M_1 = 10^{10}$ GeV,
and  the measured    
neutrino parameters to  their central values.
To mimic the possibility that 
$\beta$ and $\delta$ could be determined
$\pm 15^o$,  $\beta$-$\delta$  space is divided into  50 squares.
In each square, the programme randomly
generates values for:  $\beta, \delta $, $.001< \theta_{13} <.2$,
the smallest neutrino mass $< \sqrt{\Delta m^2_{sol}}/10$, and the
three complex angles of the $R$ matrix.
It  estimates the baryon asymmetry
from the analytic approximations of \cite{matters},
and puts a cross if it is big enough.
The programme is a proto-Monte-Carlo-Markov-Chain,
preferring to explore  parameter space where
the baryon asymmetry is large enough. 

Parametrising with the  $R$ matrix imposes
a particular measure (prior) on parameter space. This could mean we  only
explore a  class of models.
This is ok because the aim is  only to show that,  for any
PMNS phases, a large
enough  asymmetry {\it can}  be found.

{\it Discussion: } The relevant question, in discussing
the ``relation'' between CP violation in the PMNS matrix and
in leptogenesis, is whether the  baryon asymmetry
is {\it sensitive} to the PMNS phases. The answer was
``no'' for unflavoured leptogenesis in the Standard Model
seesaw\cite{Branco:2001pq}.  
This was not surprising; the seesaw contains
more phases than the PMNS matrix, and many unmeasurable
real parameters which can be ajusted to obtain a big enough asymmetry.
 In this paper,
we argue that the
answer does not change with the
inclusion of flavour effects in leptogenesis:
for any value of the PMNS phases, it is possible to
find a point in the  space of unmeasurable seesaw parameters,
such that leptogenesis works.
This ``flavoured'' asymmetry  can be written as a function of
PMNS phases, and  unmeasurables as entered
the unflavoured calculation. These can still be ajusted
to get a big enough asymmetry. 
In view of this discouraging conclusion, it is
maybe worth to emphasize that  CP violation
from  both  the left-handed and right-handed
neutrino sectors, contributes  both
to the PMNS matrix and  the baryon asymmetry.
Moreover, the answer to this question in an MSUGRA framework, 
with additional information from lepton flavour violating 
observables\cite{Borzumati:1986qx}, 
is still work in progress.

In the demonstration  that the baryon asymmetry
(produced via thermal leptogenesis) is insensitive
to PMNS phases, we found an interesting approximation
for the ``phase of leptogenesis'' (see eqn (\ref{belle})), when all lepton
flavours are in strong washout.

\section*{Acknowledgments}

J.G. is supported by a MEC-FPU Spanish grant.
This project was partially supported by
the Financement Projets Th\'eorie de l'IN2P3,
by Spanish grants FPA-2004-00996 and  FPA2005-01269
and by the EC RTN network MRTN-CT-2004-503369.


\begin{thebibliography}{222222}

%\cite{Sakharov:1967dj}
\bibitem{Sakharov:1967dj}
  A.~D.~Sakharov,
  %``Violation of CP Invariance, c Asymmetry, and Baryon Asymmetry of the
  %Universe,''
  Pisma Zh.\ Eksp.\ Teor.\ Fiz.\  {\bf 5} (1967) 32
  [JETP Lett.\  {\bf 5} (1967\ SOPUA,34,392-393.1991\ UFNAA,161,61-64.1991) 24].
  %%CITATION = UFNAA,161,NO.561;%%



\bibitem{review}
  M.~Fukugita and T.~Yanagida,
  %``Baryogenesis Without Grand Unification,''
  Phys.\ Lett.\  B {\bf 174} (1986) 45.
  %%CITATION = PHLTA,B174,45;%%
%\cite{Giudice:2003jh}
%\bibitem{Giudice:2003jh}
  G.~F.~Giudice {\it et al.},
 % A.~Notari, M.~Raidal, A.~Riotto and A.~Strumia,
  %``Towards a complete theory of thermal leptogenesis in the SM and MSSM,''
  Nucl.\ Phys.\  B {\bf 685} (2004) 89
  [arXiv:hep-ph/0310123].
  %%CITATION = NUPHA,B685,89;%%
%\cite{Buchmuller:2004nz}
%\bibitem{Buchmuller:2004nz}
  W.~Buchmuller, P.~Di Bari and M.~Plumacher,
  %``Leptogenesis for pedestrians,''
  Annals Phys.\  {\bf 315} (2005) 305
  [arXiv:hep-ph/0401240].
  %%CITATION = APNYA,315,305;%%
%\cite{Pilaftsis:2005rv}
%\bibitem{Pilaftsis:2005rv}
  A.~Pilaftsis and T.~E.~J.~Underwood,
  %``Electroweak-scale resonant leptogenesis,''
  Phys.\ Rev.\  D {\bf 72} (2005) 113001
  [arXiv:hep-ph/0506107].
  %%CITATION = PHRVA,D72,113001;%



%\cite{Branco:2001pq}
\bibitem{Branco:2001pq}
G.~C.~Branco  {\it et al.},
% T.~Morozumi, B.~M.~Nobre and M.~N.~Rebelo,
%``A bridge between CP violation at low energies and leptogenesis,''
Nucl.\ Phys.\ B {\bf 617} (2001) 475
[arXiv:hep-ph/0107164].
%%CITATION = HEP-PH 0107164;%%




\bibitem{issues}
A.~Abada  {\it et al.},
% S.~Davidson, F.~X.~Josse-Michaux, M.~Losada and A.~Riotto,
%``Flavour issues in leptogenesis,''
JCAP {\bf 0604} (2006) 004
[arXiv:hep-ph/0601083].
%%CITATION = HEP-PH 0601083;%%

\bibitem{Netal}
%\cite{Nardi:2006fx}
  E.~Nardi  {\it et al.},
  %  Y.~Nir, E.~Roulet and J.~Racker,
  %``The importance of flavor in leptogenesis,''
  JHEP {\bf 0601} (2006) 164
  [arXiv:hep-ph/0601084].
  %%CITATION = JHEPA,0601,164;%%

\bibitem{matters}
  A.~Abada  {\it et al.},
  % S.~Davidson, A.~Ibarra, F.~X.~Josse-Michaux, M.~Losada and A.~Riotto,
  %``Flavour matters in leptogenesis,''
  JHEP {\bf 0609} (2006) 010
  [arXiv:hep-ph/0605281].
  %%CITATION = JHEPA,0609,010;%%


%\cite{Fujihara:2005pv}
\bibitem{Fujihara:2005pv}
  T.~Fujihara  {\it et al.},
  % S.~Kaneko, S.~Kang, D.~Kimura, T.~Morozumi and M.~Tanimoto,
  %``Cosmological family asymmetry and CP violation,''
  Phys.\ Rev.\  D {\bf 72} (2005) 016006
  [arXiv:hep-ph/0505076].
  %%CITATION = PHRVA,D72,016006;%%
%
\bibitem{Pascoli:2006ie}
  S.~Pascoli, S.~T.~Petcov and A.~Riotto,
  %``Connecting low energy leptonic CP-violation to leptogenesis,''
  Phys.\ Rev.\  D {\bf 75} (2007) 083511
  [arXiv:hep-ph/0609125].
  %%CITATION = PHRVA,D75,083511;%%

%\cite{Branco:2006ce}
\bibitem{Branco:2006ce}
  G.~C.~Branco, R.~Gonzalez Felipe and F.~R.~Joaquim,
  %``A new bridge between leptonic CP violation and leptogenesis,''
  Phys.\ Lett.\  B {\bf 645} (2007) 432
  [arXiv:hep-ph/0609297].
  %%CITATION = PHLTA,B645,432;%%

\bibitem{akr}
 S.~Antusch, S.~F.~King and A.~Riotto,
  %``Flavour-dependent leptogenesis with sequential dominance,''
  JCAP {\bf 0611} (2006) 011
  [arXiv:hep-ph/0609038].


\bibitem{seesaw} 
P.~Minkowski,
%``Mu $\to$ E Gamma At A Rate Of One Out Of 1-Billion Muon Decays?,''
Phys.\ Lett.\ B {\bf 67} (1977) 421;
%%CITATION = PHLTA,B67,421;%%
M. Gell-Mann, P. Ramond and
R. Slansky,  {\em Proceedings of the Supergravity Stony Brook Workshop}, New
York 1979,  eds. P. Van Nieuwenhuizen and D. Freedman; T. Yanagida,  {\em
Proceedinds of the Workshop on Unified Theories and Baryon Number in the
Universe},  Tsukuba, Japan 1979, ed.s A. Sawada and A. Sugamoto;
R. N. Mohapatra, G. Senjanovic,
{\it Phys.Rev.Lett.} {\bf 44} (1980)912.
%


\bibitem{ss25}
S.~Davidson,
%``Parametrizations of the seesaw, or, can the seesaw be tested?,''
arXiv:hep-ph/0409339.
%%CITATION = HEP-PH 0409339;%%

%\cite{Blanchet:2006ch}
\bibitem{BdBR}
  S.~Blanchet, P.~Di Bari and G.~G.~Raffelt,
  %``Quantum Zeno effect and the impact of flavor in leptogenesis,''
  arXiv:hep-ph/0611337.
  %%CITATION = HEP-PH/0611337;%%



%\cite{Barbieri:1999ma}
\bibitem{Barbieri}
  R.~Barbieri {\it et al.},
  % P.~Creminelli, A.~Strumia and N.~Tetradis,
  %``Baryogenesis through leptogenesis,''
  Nucl.\ Phys.\  B {\bf 575} (2000) 61
  [arXiv:hep-ph/9911315].
  %%CITATION = NUPHA,B575,61;%%


\bibitem{AAFX}
%\cite{Josse-Michaux:2007zj}
%\bibitem{Josse-Michaux:2007zj}
  F.~X.~Josse-Michaux and A.~Abada,
  %``Study of flavour dependencies in leptogenesis,''
  arXiv:hep-ph/0703084.
  %%CITATION = HEP-PH/0703084;%%


%\cite{Hamaguchi:2001gw}
\bibitem{HMY}
  K.~Hamaguchi, H.~Murayama and T.~Yanagida,
  %``Leptogenesis from sneutrino-dominated early universe,''
  Phys.\ Rev.\  D {\bf 65} (2002) 043512
  [arXiv:hep-ph/0109030].
  %%CITATION = PHRVA,D65,043512;%%



%\cite{Casas:2001sr}
\bibitem{Casas:2001sr}
  J.~A.~Casas and A.~Ibarra,
  %``Oscillating neutrinos and mu --> e, gamma,''
  Nucl.\ Phys.\  B {\bf 618} (2001) 171
  [arXiv:hep-ph/0103065].
  %%CITATION = NUPHA,B618,171;%%






%\cite{Davidson:2003yk}
%\bibitem{Davidson:2003yk}
%S.~Davidson and R.~Kitano,
%%``Leptogenesis and a Jarlskog invariant,''
%JHEP {\bf 0403} (2004) 020
%[arXiv:hep-ph/0312007].
%%CITATION = HEP-PH 0312007;%%

%\cite{Davidson:2003cq}
%\bibitem{DIP}
%S.~Davidson,
%``From weak-scale observables to leptogenesis,''
%JHEP {\bf 0303} (2003) 037
%[arXiv:hep-ph/0302075].
%%CITATION = HEP-PH 0302075;%%




%\cite{Spergel:2006hy}
\bibitem{Spergel:2006hy}
  D.~N.~Spergel {\it et al.}  [WMAP Collaboration],
  %``Wilkinson Microwave Anisotropy Probe (WMAP) three year results:
  %Implications for cosmology,''
  arXiv:astro-ph/0603449.
  %%CITATION = ASTRO-PH/0603449;%%



%\cite{Borzumati:1986qx}
\bibitem{Borzumati:1986qx}
  F.~Borzumati and A.~Masiero,
  %``Large Muon And Electron Number Violations In Supergravity Theories,''
  Phys.\ Rev.\ Lett.\  {\bf 57} (1986) 961.
  %%CITATION = PRLTA,57,961;%%
J.~Hisano {\it et al.},
  % T.~Moroi, K.~Tobe and M.~Yamaguchi,
  %``Lepton-Flavor Violation via Right-Handed Neutrino Yukawa Couplings in
  %Supersymmetric Standard Model,''
  Phys.\ Rev.\  D {\bf 53} (1996) 2442
  [arXiv:hep-ph/9510309].
  %%CITATION = PHRVA,D53,2442;%%
Y.~Kuno and Y.~Okada,
  %``Muon decay and physics beyond the standard model,''
  Rev.\ Mod.\ Phys.\  {\bf 73} (2001) 151
  [arXiv:hep-ph/9909265].
  %%CITATION = RMPHA,73,151;%%

\end{thebibliography}
\end{document}